\documentclass[letterpaper, 10 pt, conference]{ieeeconf}  
\IEEEoverridecommandlockouts                              
\usepackage{epsfig} 
\usepackage{graphicx}
\usepackage{amsmath}
\usepackage{array}
\usepackage{amssymb}
\usepackage{verbatim}
\usepackage{cleveref}
\usepackage{color}
\usepackage{relsize}
\usepackage{tikz}
\usepackage{fancyhdr}
\usepackage{graphics}
\usepackage{setspace}
\usepackage{epstopdf}
\usepackage{multirow}
\usepackage{lipsum}
\usepackage{textcomp}
\usepackage{mathtools}
\usepackage{cite}
\usepackage{setspace}
\usepackage{booktabs}
\usepackage{breqn}
\usepackage{algorithm}
\usepackage[noend]{algpseudocode}
\usepackage{flushend}
\usepackage{multicol}
\usepackage{float}
\usepackage{mathtools,bm}
\usepackage[noend]{algpseudocode}
\usepackage{booktabs}
\usepackage{dblfloatfix}
\usepackage{mathrsfs}
\usepackage{mathtools}

\fancypagestyle{firstpage}{\lhead{\vspace{-0.75 cm} \small \textbf{{\fontfamily{cmss}\selectfont
The 3rd IEEE Conference on Control Technology and Applications (CCTA)\\August 19--21, 2019, Hong Kong, China}}}}

\title{\large \bf
Thermal Responses of Connected HEVs Engine and Aftertreatment Systems to Eco-Driving$^*$}

\author{Mohammad Reza Amini$^{1}$, Yiheng Feng$^{2}$, Hao Wang$^{1}$, Ilya Kolmanovsky$^{3}$, and Jing Sun$^{1}$
\thanks{*This paper is based upon the work supported by the United States DoE, ARPA-E under award No. DE-AR0000797.}
\thanks{$^{1}$M.R. Amini, H. Wang, and J. Sun are with the Dept. of Naval Architecture \& Marine Engineering, University of Michigan, Ann Arbor, MI 48109 USA. Emails: {\tt\small \{mamini,autowang,jingsun\}@umich.edu}}%
\thanks{$^{2}$Y. Feng is with the Transportation Research Institute, University of Michigan, Ann Arbor, MI, 48109, USA. Email: {\tt\small yhfeng@umich.edu}
}
\thanks{$^{3}$I. Kolmanovsky is with  the Dept. of Aerospace Engineering, University of Michigan, Ann Arbor, MI 48109 USA. Email: {\tt\small ilya@umich.edu}}%
}

\renewcommand{\baselinestretch}{1}
\begin{document}

\maketitle
\thispagestyle{firstpage}

\begin{abstract}
Connected and automated vehicles (CAVs) have been recognized as providing unprecedented opportunities for substantial fuel economy improvement through CAV-based vehicle speed trajectory optimization (eco-driving). At the same time, the implications of the CAV operation on thermal responses, including those of engine and exhaust aftertreatment system, have not been fully investigated. To this end, firstly, a sequential optimization framework for vehicle speed trajectory planning and powertrain control in hybrid electric CAVs is proposed in this paper. Next, the impact of eco-driving and power split optimization on the engine and catalytic converter thermal responses, as well as on the tailpipe emissions is characterized. Despite an average 16\% improvement in fuel economy through sequential optimization, this study shows that eco-driving slows down the thermal responses, which could unfavorably affect the tailpipe emissions.
 \end{abstract}

\vspace{-0.25cm}
\section{INTRODUCTION}
The research into connected and automated vehicles (CAVs), equipped with advanced sensors and capable of communication with infrastructure (V2I) and surrounding vehicles (V2V), has been extended in recent years to areas beyond the original intents of enhancing safety and improving mobility\cite{sciarretta2015optimal}. Connectivity enables more accurate prediction of future traffic conditions so that the vehicle speed profile can be optimized for eco-driving and fuel/energy saving~\cite{prakash2016assessing,barth2009energy}. Over the recent years, extensive studies have been carried out on eco-driving and powertrain optimization for CAVs (see e.g.,~\cite{guanetti2018control}, and the references therein). Most of the recent CAV-related research, however, has been focused on reducing losses related to traction, while little has been reported on the impact on the thermal responses of the CAVs. 

Thermal management is a significant factor in the overall vehicle fuel economy, especially for electrified vehicles~\cite{AminiCDC18}. In electrified vehicles, the battery is used to meet the power demand for traction, auxiliary systems that include the cooling and heating of the cabin (i.e., HVAC), and thermal management for engine and battery in the case of HEVs. For light-duty HEVs, air conditioning (A/C) of the passenger compartment is the most significant auxiliary load on the battery~\cite{Rugh2008}. Efficient thermal management of the engine is also required for efficient operation of the engine and of its aftertreatment system~\cite{donkers2017optimal}. In particular, the efficiency of the catalytic converter and the tailpipe emission levels are temperature-dependant. Given the close coupling between the power and thermal loops of an electrified vehicle, optimizing the power system through eco-driving will affect the thermal responses. Thus, understanding this impact is important for the design and optimization of eco-driving strategy.

This paper focuses on investigating the impact of eco-driving and power split optimization of connected HEVs on the thermal responses of the engine and exhaust aftertreatment systems. To this end, we propose a sequential optimization framework to assess the benefits of using traffic information for optimizing the vehicle speed and power split for an HEV operating in a CAV driving environment. The sequential optimization framework has two stages: (i) planning the optimal speed trajectory using traffic information, and (ii) seeking optimal power split between the engine and the battery using dynamic programming (DP). An eco-driving approach developed in our previous work~\cite{Zhen2018Traj} for power-split HEV in congested urban traffic environment is considered. The vehicle queuing process is explicitly modeled by the shockwave profile model (SPM)~\cite{wu2011shockwave} to provide a green window for eco-vehicle speed trajectory planning. Next, using experimentally validated thermal models, the impact of sequential optimization on the thermal responses, including engine coolant temperature, catalyst temperature, and tailpipe emissions, is investigated. Our results support the need for an integrated power and thermal management (iPTM) approach for CAVs and provide tools and guidelines for an effective trade-off between fuel economy and emission levels within an iPTM optimization framework.

\vspace{-0.2cm} 
\section{Traffic Modeling}\label{sec:Sec2}\vspace{-0.05cm} 
{Surrounding traffic plays an important role in determining vehicle trajectories and therefore fuel consumption, especially when the vehicle is approaching a signalized intersection. Through V2I communication, the infrastructure is able to receive real-time vehicle status including location and speed of all surrounding vehicles. With traffic signal information, the infrastructure is able to predict vehicle queuing dynamics at the intersection and generate a green time window for eco-driving trajectory planning~\cite{Zhen2018Traj}.}
 
{In this paper, the queuing process is modeled based on the SPM which tracks different types of shockwaves to estimate critical time points for green window calculation. The green window is defined as the time interval during which an eco-driving vehicle can pass through the intersection. In this study, prediction of the queuing dynamics before the eco-driving vehicle's arrival at the intersection is required for vehicle speed planning. The SPM is therefore modified to represent vehicle acceleration and deceleration process and make predictions of the queuing dynamics. 
The details of queuing process modelling using the SPM can be found in our previous works~\cite{Zhen2018Traj,amini2019sequential}.}

{To illustrate our approach to traffic modeling, a six-intersection corridor on Plymouth Road in Ann Arbor (as in~\cite{amini2019sequential}) is modeled. The stretch of the road represented in the simulations is about 2.2 miles and has two lanes for each direction. A microscopic traffic simulation software VISSIM~\cite{ptv2016ptv} is used to build the road network and simulate background traffic. To calibrate the simulation model and represent congested traffic, real-world data were collected during PM rush hour (4:00 PM-5:00 PM), including traffic volume, turning ratio, and traffic signal timing at each intersection. Vehicles in VISSIM are programmed to broadcast basic safety messages (BSMs). Additionally, all the traffic signals are programmed to broadcast signal phasing and timing (SPaT) in real-time. All vehicle and traffic signal data are sent to the queuing profile algorithm for prediction, as described in~\cite{amini2019sequential}. Finally, the predicted green window is sent to the trajectory planning algorithm, which will be discussed in Sec.~\ref{sec:sec4}.}\vspace{-0.1cm} 
%

\section{HEV Power and Thermal Models}
\begin{figure}[h!]
	\begin{center}
		\includegraphics[width=0.91\columnwidth]{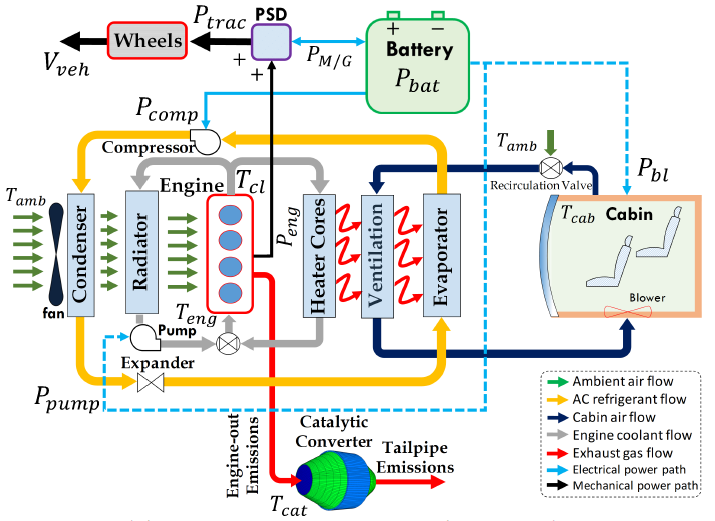} \vspace{-0.4cm}   
		\caption{Schematic of power split HEV thermal and power loops.}\vspace{-0.9cm} \label{fig:HEV_PS_AC_Schematic} 
	\end{center}
\end{figure}

The overall schematic of a power split HEV with thermal and power loops is shown in Fig.~\ref{fig:HEV_PS_AC_Schematic}. The demanded traction power $P_{trac}$ is provided by the combustion engine and the electric battery. The ratio of the engine power ($P_{eng}$) to the battery power provided for traction (and consumed by two motor/generators ($M/G$)) is determined by power management system with the power split device (PSD) as the actuator. In addition to the electric propulsion power ($P_{M/G}$), the battery provides power for auxiliary loads. In this paper, we consider the major auxiliary power loads for the A/C compressor ($P_{comp}$), HVAC blower ($P_{bl}$), and the electric pump of the engine coolant ($P_{pump}$), namely, $P_{aux}=P_{comp}+P_{bl}+P_{pump}$. Fig.~\ref{fig:HEV_PS_AC_Schematic} also shows the cabin cooling and heating loops, in which the A/C compressor is the primary actuator for cooling the cabin, while the waste heat of the engine is used to heat the cabin via the engine coolant circulation.~
Another thermal subsystem of the HEV is the aftertreatment system equipped with a three-way catalytic converter (TWC). The conversion efficiency of the TWC is dictated by the catalyst temperature ($T_{cat}$), which is a function of the engine-out exhaust gas flow and temperature.
\vspace{-0.45cm}

\vspace{-0.15cm}   
\subsection{Battery Power-Balance Model}
A simplified phenomenological HEV powertrain model based on the mechanical and electrical power balance is adopted from our previous work~\cite{amini2019sequential}. The response of the battery state of charge ($SOC$) is the main dynamics captured by the model, based on which the power split between the engine and the battery is determined by the powertrain controller. The model, which captures the impact of the traction ($P_{M/G}$) and auxiliary ($P_{aux}$) power loads, has the following structure~\cite{amini2019sequential}: \vspace{-0.2cm}   
%
\begin{gather}\label{eq:SOC_simple_model}
SOC(k+1)=SOC(k)+~~~~~~~~~~~~~~~~~~~~~~~~~~~~~~~\\
\begin{cases}
\xi_1 P_{M/G}(k)+\xi_2 P_{M/G}^2(k)+\xi_3 P_{M/G}(k)P_{aux}(k)\\
~~~~~+\xi_4 P_{aux}(k)+\xi_5 P_{aux}^2(k)+\xi_6,~~~~~\text{if A/C On;}\\
~\vspace{-0.3cm}\\
\xi_7 P_{M/G}(k)+\xi_8 P_{M/G}^2(k)+\xi_9,~~~~~~~~~~~\text{if A/C Off.} 
\end{cases}\nonumber
\end{gather}
\normalsize
The $SOC$ model in~(\ref{eq:SOC_simple_model}) has a switching structure depending on whether the A/C system is on or off. The main reason for selecting the switching structure for (\ref{eq:SOC_simple_model}) is the considerable power consumed by the A/C system (e.g., up to 2.5 $kW$). When the A/C is off, $P_{aux}$ is negligible and the $SOC$ dynamics can be captured with $P_{M/G}$ as the input. Parameters $\xi_{1,\cdots,9}$ in (\ref{eq:SOC_simple_model}) are identified from experimental data collected from our test vehicle (2017 MY Prius) with sampling time $T_s=1~sec$, see~\cite{amini2019sequential} for more details on the experimental validation and the identified parameters of the model in~(\ref{eq:SOC_simple_model}). 

\vspace{-0.2cm} 
\subsection{Engine Coolant Thermal Model}
\vspace{-0.05cm} 
The heat transferred from the engine to the coolant is normally dissipated into the ambient environment at the radiator (see Fig.~\ref{fig:HEV_PS_AC_Schematic}). When there is cabin heating demand, a part of the thermal energy stored in the coolant is transferred to the cabin at the heater cores. The dynamics of the engine coolant temperature~\cite{kim2015control} sampled at $T_s=1~sec$ is modeled as follows~\cite{XunCCTA19}:
\vspace{-0.25cm} 
\begin{gather}
\label{eq:HEVmodel:dotTcl}
{T}_{cl}(k+1)=T_{cl}(k)+\frac{1}{M_{e}C_{e}}(\dot{Q}_{fuel}(k)-P_{eng}(k)\\-\dot{Q}_{exh}(k)-\dot{Q}_{air}(k)-\dot{Q}_{rad}(k)-\dot{Q}_{heat}(k)),\nonumber
\end{gather}
where $M_{e}$, $C_{e}$ and $T_{cl}$ are the mass, the equivalent specific heat capacity, and the coolant temperature of the engine respectively, $\dot{Q}_{fuel}$ is the heat released in the combustion of the fuel, $\dot{Q}_{exh}$ is the heat rejected by exhaust gas, $\dot{Q}_{air}$ is the heat rejected by the convection from the engine to the air, $\dot{Q}_{rad}$ is the heat rejected by radiator/fan, and $\dot{Q}_{heat}$ is the heating power delivered to the cabin. In this paper, we assume there is no heating requirement for the cabin, as we focus on the operation of the vehicle in the summer, during which the A/C system is used for cooling the cabin air. The model (\ref{eq:HEVmodel:dotTcl}) has been experimentally validated using the data collected from the test vehicle over highway and city driving cycles in Ann Arbor, MI, and the results are reported in~\cite{XunCCTA19}. 
%

\vspace{-0.15cm} 
\subsection{Aftertreatment Thermal Model}
\begin{figure}[h!]
	\begin{center}
		\includegraphics[width=0.8\columnwidth]{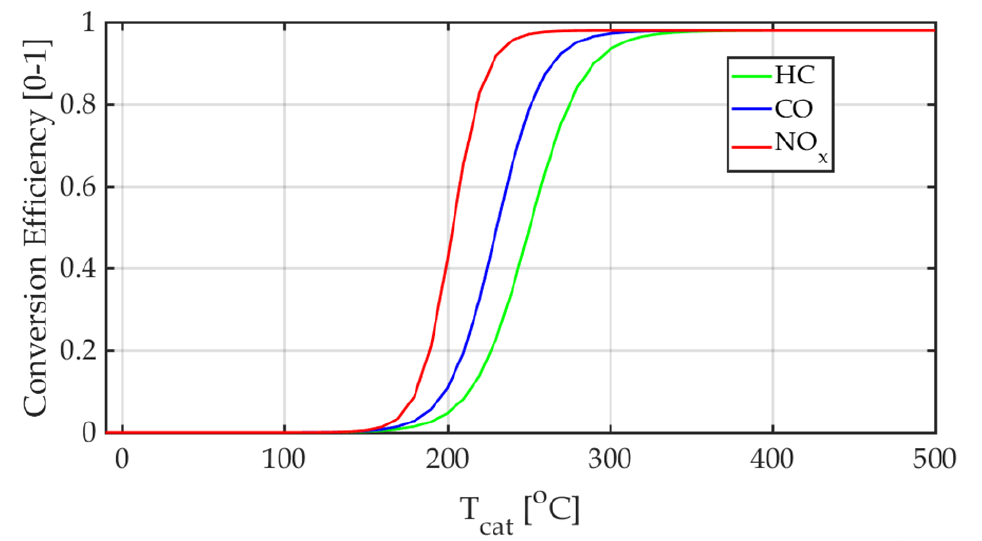} \vspace{-0.35cm}   
		\caption{Conversion efficiencies of the TWC for HC, NOx, and CO as functions of the catalyst temperature $T_{cat}$.}\vspace{-0.55cm} 
		\label{fig:ATS_Eff_Curves} 
	\end{center}
\end{figure}

\vspace{-0.05cm} 
A physics-based model of the three-way catalytic converter (TWC) from ADVISOR~\cite{markel2002advisor} is considered in this paper to evaluate the eco-driving and power split optimization impact on the tailpipe emissions. This model captures the catalyst temperature dynamics ($T_{cat}$) and uses static maps for estimating the engine-out emissions, including carbon monoxide (CO), unburned hydrocarbon (HC), and nitrogen oxide (NOx). The ADVISOR model has been parameterized to fit $T_{cat}$ data collected from our test vehicle. See~\cite{markel2002advisor} for more details on the TWC thermal model. Catalyst temperature $T_{cat}$ is the main state of the TWC, as the conversion efficiency of catalytic converters is a strong function of $T_{cat}$. Fig.~\ref{fig:ATS_Eff_Curves} shows the conversion efficiency of the TWC for HC, CO, and NOx as a function of $T_{cat}$. It can be seen from Fig.~\ref{fig:ATS_Eff_Curves} that the catalyst does not ``light-off'' (i.e., does not reach the conversion efficiency of $0.5$) until its temperature reaches about $200^oC$.
~Additionally, it can be observed that compared to NOx and CO, the light-off for the HC occurs at a relatively higher temperature, e.g., 250$^oC$~\cite{shahbakhti2015early}. \vspace{-0.3cm} 

\vspace{-0.25cm} 
\section{Sequential Optimization of Speed \\and Power Split}\label{sec:sec4} \vspace{-0.1cm} 
In this section, two stages of the sequential eco-driving and power split optimization are described. In the first stage, the vehicle speed is optimized according to the predicted queue. The outputs from the first optimization stage are the optimized vehicle speed and traction power trajectories. In this paper, we assume a constant auxiliary load for the HVAC system of $P_{aux}=1700W$ (average $P_{aux}$ observed in the test vehicle). The constant $P_{aux}$ along with the optimized traction power from the first stage are used in the second optimization stage, where the optimal power split between mechanical engine power and battery electrical power, as well as the engine operating mode are determined. \vspace{-0.1cm}  

\subsection{\textbf{Stage I}: Vehicle Speed Optimization (Eco-Driving)}
%
{Given the green window (described in Sec.~\ref{sec:Sec2} and~\cite{Zhen2018Traj,amini2019sequential}) which prescribes a time interval for the eco-driving vehicle to arrive at the intersection, the speed optimization algorithm generates an eco-friendly vehicle trajectory. The planning horizon of the trajectory starts from the time instant the eco-driving vehicle enters the communication range until the vehicle departs from the intersection. In order to ensure a smooth trajectory and reduce fuel consumption, a trigonometric speed profile from~\cite{barth2011dynamic} is applied, see~\cite{amini2019sequential,barth2011dynamic} for the details.}
%
%

{Based on the relationship between the predicted green window, traffic signal status, and the remaining time, the eco-driving vehicle may choose one of the following four speed profiles: ``slow down'', ``speed up'', ``cruise" or ``stop". All speed profiles except for ``cruise'' are based on the trigonometric profiles while the ``cruise'' speed profile maintains a constant speed to pass the intersection. In some of the previous studies, the ``stop'' speed profile is ignored because the eco-driving vehicle can always slow down to a very low speed and pass the intersection without a stop. However, a very low cruise speed may obstruct other vehicles and cause frequent lane changing and cut-in behaviors. As a result, the minimum cruise speed is set to be 70\% of the speed limit. If the eco-driving vehicle cannot meet this speed limit, it will choose the ``stop" speed profile, see~\cite{Zhen2018Traj}.\vspace{-0.10cm} 

\subsection{\textbf{Stage II}: Power Split Optimization}
In the second optimization stage, the split between mechanical (engine) and electrical (battery) power is optimized by using dynamic programming (DP). The optimized vehicle speed (traction power) from Stage I and $P_{aux}$ are used as known inputs, and the objective is to find the optimum battery power and engine operating mode to minimize the overall energy consumption of the HEV while enforcing the vehicle operating constraints. With a total travel time of $\mathcal{T}$ discretized by $K$ sampling instants, the most fuel efficient operation can be obtained by minimizing the following cost function:\vspace{-0.22cm}
\begin{gather}
\label{eq:Task4_DPformulation}
~~~{m_f(x,u,K)=\sum_{i=0}^K W_f(x(i),u(i)) +\Phi(x(K))},\\
\text{s.t.}~~~~~~~~~~~x(k+1) = \text{Eq.~(\ref{eq:SOC_simple_model})},~x\in \mathcal{X},~u \in \mathcal{U},
\end{gather}
\normalsize
where $x=SOC$ is the state, the control input $u$ includes the engine operation mode and battery power, i.e., $u=[e_{mode}, P_{bat}]^{\intercal}$, and $\Phi(x(K))$ represents the terminal cost on $SOC$. Note that $P_{bat}=P_{M/G}+P_{aux}$, where $P_{M/G}$ and $P_{aux}$ are the inputs to $SOC$ model (\ref{eq:SOC_simple_model}). Moreover, the following constraints should be satisfied in different operating modes:
\begin{itemize}
\item Engine Off:~$e_{mode}=1$,~$P_{eng}=0,~\omega_e=0,~W_f=0$,
\item Engine On:~$e_{mode}=2$,~$P_{eng}>0,~\omega_e\neq 0,~W_f=f(\omega_{eng},P_{eng})$,
\end{itemize}
%
where $\omega_{eng}$ and $P_{eng}$ are the speed and output power of the engine, respectively. The engine fuel consumption is given by a function of the engine speed and power, $f(\omega_{eng},P_{eng})$. This map has been developed from experimental data, see~\cite{XunCCTA19} for the details. When $e_{mode}=2$, it is assumed that the engine operates on the optimal operation line~\cite{XunCCTA19}.  \vspace{-0.15cm}
 
\section{Simulation Results and Discussions}\vspace{-0.05cm}
In this section, the results of sequential optimization are presented and compared to the rule-based algorithm, which was tuned to provide charge sustaining performance. The overall equivalent vehicle energy consumption calculation accounts for both fuel and $SOC$ change during the driving cycle.

 \vspace{-0.15cm}
\subsection{\textbf{Impact on Fuel Economy}}
Fifty vehicles are simulated using different random initial speeds sampled from a uniform distribution ranging from 48 $km/h$ to 58 $km/h$ and they proceed through the same corridor modeled in VISSIM. Fig.~\ref{fig:AllCases_SpeedOptz_EnergyComparison_Percent} shows the energy saving results achieved by (i) speed optimization when the rule-based controller is considered for power split, and (ii) speed optimization when the power split is optimized via DP. Compared with the non-optimized speed case with rule-based power split controller (i.e., the \textit{baseline} case), it can be seen from Fig.~\ref{fig:AllCases_SpeedOptz_EnergyComparison_Percent} that the overall energy consumption for speed optimization alone is reduced by 13\% on average, with the maximum reduction as high as 28.6\%. When the power split optimization is included, the average energy saving increases from 13\% to 16\%. It can be seen that for two cases out of fifty, the optimized speed profiles with rule-based power split controller result in slightly higher energy consumption (0.15\%). It is possible that in these two cases the speed trajectory planning is more conservative, while the vehicle with non-optimized speed may pass the intersection when the signal turns to yellow. 
\vspace{-0.35cm}   
\begin{figure}[h!]
 	\begin{center}
 		\includegraphics[width=0.91\columnwidth]{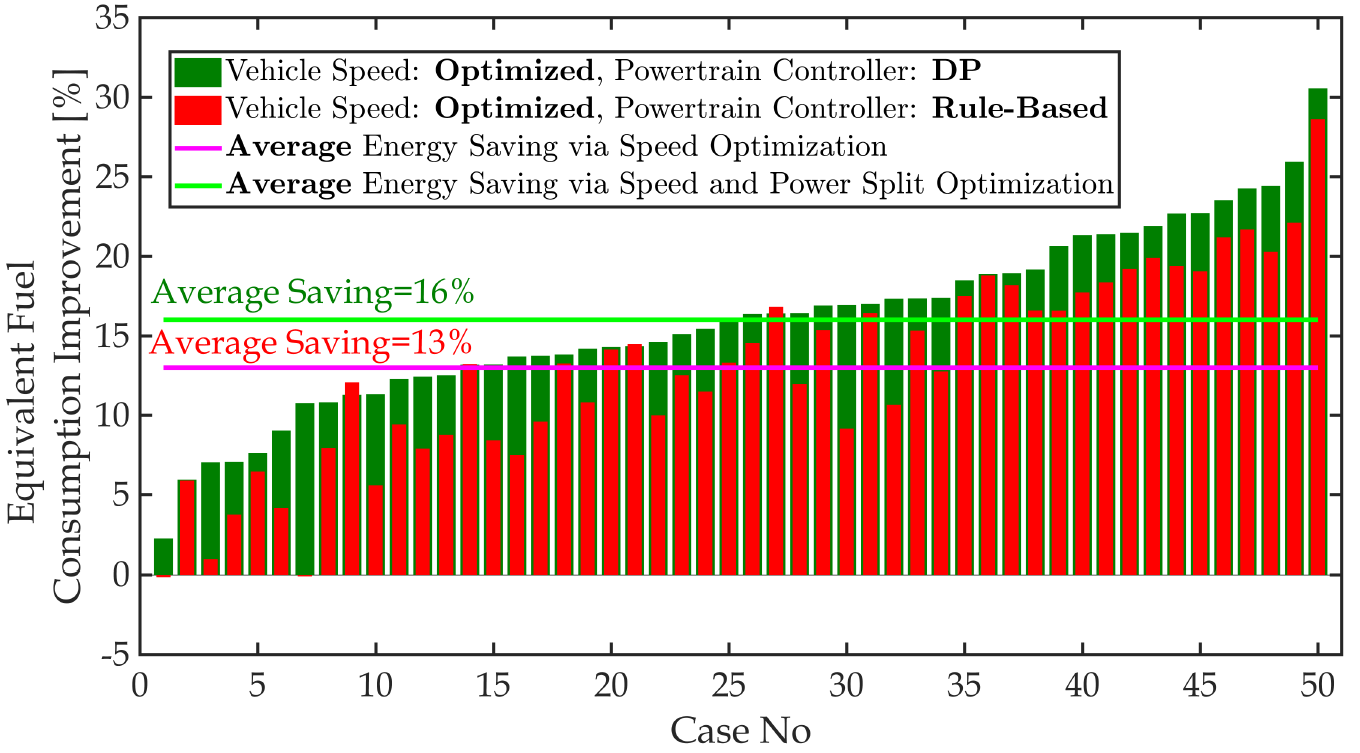} \vspace{-0.4cm} 
 		\caption{Energy saving results via speed and power split optimization for 50 cases (vehicles) simulated over the same route modeled in VISSIM~\cite{amini2019sequential}.}\vspace{-0.55cm} \label{fig:AllCases_SpeedOptz_EnergyComparison_Percent} 
 	\end{center}
\end{figure}
%

%

For better understanding of the impact of speed and power split optimization on the HEV powertrain performance, the trajectories of the powertrain system for one of the simulated cases are plotted in Fig.~\ref{fig:DP_SpeedOptimization_SOCEng_Nov28_2}. For the selected case, 11.9\% (Stage I) and 17\% (Stage I+II) reductions in energy consumption were observed. By looking into the engine speed trajectories in Fig.~\ref{fig:DP_SpeedOptimization_SOCEng_Nov28_2}-$b$, one can see the effect of speed optimization on the engine performance. When the speed is not optimized, the engine is utilized more aggressively during the vehicle accelerations with non-smooth engine speed trajectories. Speed optimization, on the other hand, results in smoother operation of the engine by reducing the aggressiveness in the vehicle speed profile and the acceleration demand. 

The most noteworthy difference between the DP and the rule-based strategy is when the vehicle stops. During the stops, while there is no traction power demand, the auxiliary power needs to be provided. The rule-based controller turns off the engine and uses the battery to provide the auxiliary power. The DP-based solution, however, commands the battery to be charged during the vehicle stops by running the engine at low (optimal) speed. This results in more electrical power be used for propulsion during the driving phase with the energy saved in the battery while the vehicle is stopped. This results in less usage of the engine for traction during the drive, thereby lowering the overall fuel and energy consumption compared to the baseline case. DP suggests to use the engine during the long vehicle stops because it realizes that (i) the A/C system draws power from the battery all the time and the battery needs to be charged, (ii) the considered city driving cycle does not include any highway driving segment; thus, the electrical power (i.e., EV mode) could be sufficient to drive the vehicle if enough energy was saved in the battery during the vehicle stops, and (iii) running the engine at lower (optimal) speeds leads to overall higher fuel economy as compared to the rule-based controller which only uses the engine at higher speeds during the drive.
%
\begin{figure}[t!]
	\begin{center}
		\includegraphics[width=0.85\columnwidth]{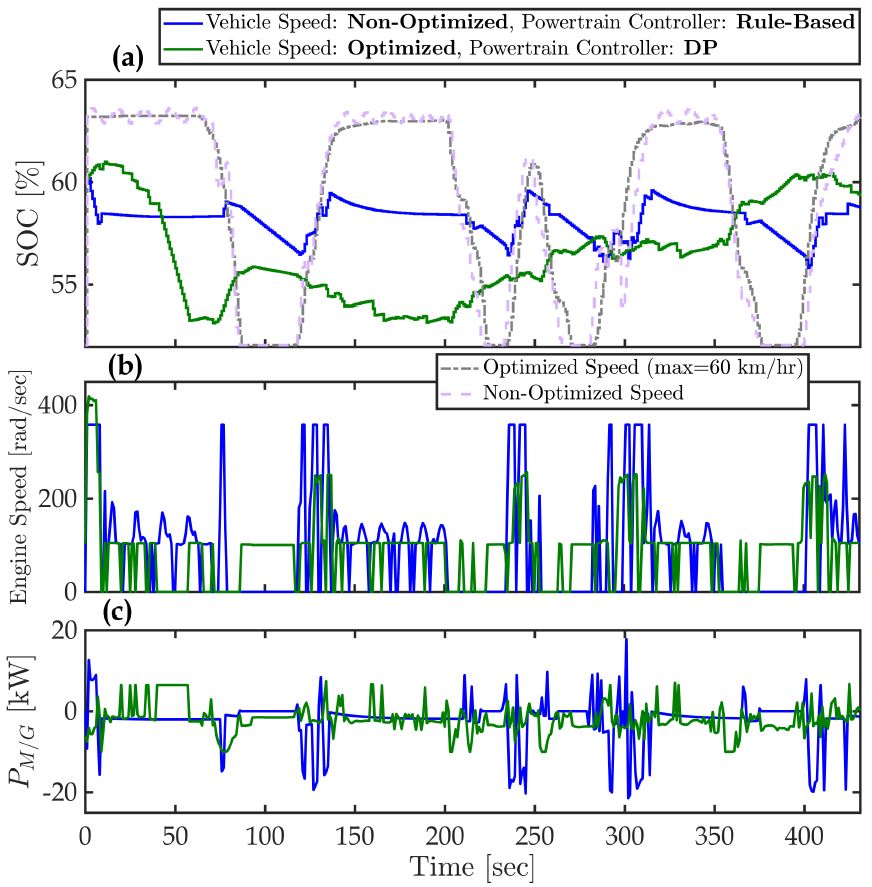} \vspace{-0.45cm}   
		\caption{The results of vehicle speed and power split optimization and its effects on vehicle powertrain performance compared to the baseline case: (a) battery $SOC$ with vehicle speed superimposed, (b) engine speed, and (c) battery electric propulsion.}\vspace{-0.85cm} 
		\label{fig:DP_SpeedOptimization_SOCEng_Nov28_2} 
	\end{center}
\end{figure}

\vspace{-0.15cm} 
\subsection{\textbf{Impact on Thermal Responses}}
%
In this section, we evaluate the impact of sequential vehicle speed and power split optimization on the thermal responses of the engine and aftertreatment system. Since we only consider the operation of the vehicle in warm/hot weather conditions, we assume that the power split logic is not affected by the coolant and catalyst temperatures. This allows for explicit investigation of the sequential optimization impact on the thermal responses. In cold weather conditions, however, the power split could be affected by the TWC and coolant temperatures~\cite{shahbakhti2015early}. Here we assume that the engine is warmed-up, e.g., $T_{cl}>50^oC$. The aftertreatment system is directly affected by the coolant temperature, as the coolant temperature affects the engine exhaust temperature. Since $SOC$ depends on the battery power which in turn depends on the power split ratio, $P_{aux}$, and the engine output power, $P_{eng}$, the dynamics of $SOC$ could be linked to $T_{cl}$ and $T_{cat}$.
%
%

  %

\subsubsection{\textbf{Impact on Coolant and Catalyst Temperatures}}Fig.~\ref{fig:ATS_coldstart_Tcl_Tcat} shows the overall thermal responses of the engine ($T_{cl}$) and TWC ($T_{cat}$) for different vehicle speed and power split control scenarios. Here, it is assumed that the engine is warmed-up, but the initial $T_{cat}=50^oC$ temperature is below the light-off temperature of the TWC. Moreover, it is assumed that the ambient temperature is constant $T_{amb}=30^oC$, and $P_{aux}=1700W$. When the power split controller is rule-based, speed optimization results in lower catalyst and engine coolant temperatures on average (Figs.~\ref{fig:ATS_coldstart_Tcl_Tcat}-$a,b$). This can be attributed to the effect of speed optimization on powertrain operation. Specifically, reducing the engine speed excursions (hikes) reduces the generated heat by the engine and the coolant temperature. Since $T_{cat}$ is affected by the engine thermal response ($T_{cl}$), $T_{cat}$ is also lower on average when the speed is optimized. Thus, compared to non-optimized speed cases, the TWC reaches the light-off temperature in a longer time. Note that this analysis does not account for special TWC light-off strategies which may be used when TWC is cold. \vspace{-0.25cm} 
\begin{figure}[h!]
 	\begin{center}
 		\includegraphics[width=1.0\columnwidth]{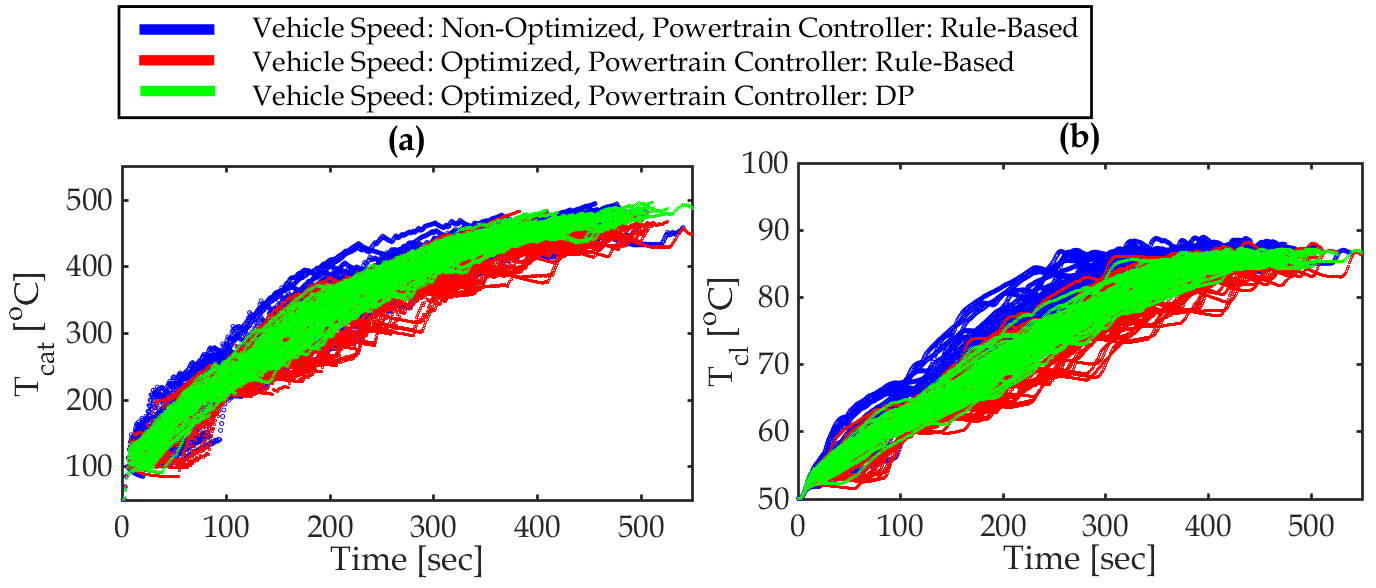} \vspace{-0.85cm}   
 		\caption{The overall thermal responses of ($a$) engine ($T_{cl}$) and ($b$) TWC ($T_{cat}$) for the fifty cases.}\vspace{-0.6cm} 
 		\label{fig:ATS_coldstart_Tcl_Tcat} 
 	\end{center}
 \end{figure}
 
  \begin{figure*}[h!]
	\begin{center}
		\includegraphics[width=1.46\columnwidth]{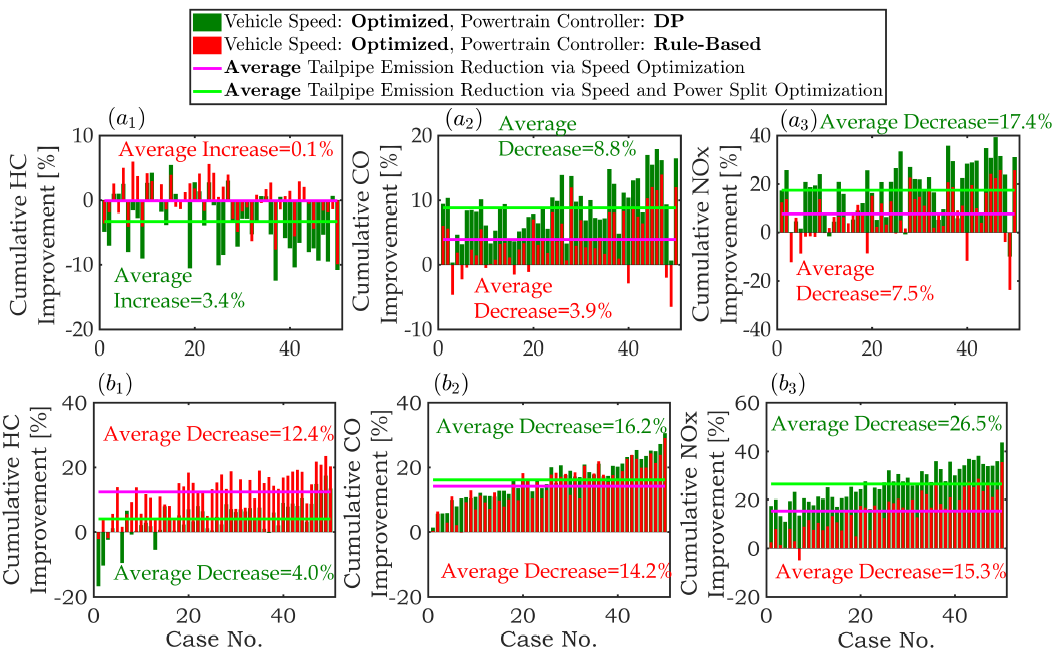}
		\vspace{-0.4cm}   
		\caption{The overall impact of sequential optimization on the tailpipe emissions: ($a$) $T_{cat,0}=50^oC$, and ($b$) $T_{cat,0}=300^oC$. The order of the fifty cases is the same as the one shown in Fig.~\ref{fig:AllCases_SpeedOptz_EnergyComparison_Percent}, i.e., arranged in the order of increasing energy savings with sequential optimization.}\vspace{-0.8cm} 
		\label{fig:Experimental_ID_Vrf_Data} 
	\end{center}
\end{figure*}

When the power split is determined by DP, as discussed in the previous section, the engine may be used at the stops to save the energy in the battery and reduce the future engine utilization during the drive. As a result, with the same optimized speed trajectories, the DP-based power split leads to higher $T_{cl}$ and $T_{cat}$. However, as shown in Fig.~\ref{fig:ATS_coldstart_Tcl_Tcat}, the average $T_{cl}$ and $T_{cat}$ are still lower than in the non-optimized speed cases with the rule-based power split controller.

\vspace{0.02cm}
\subsubsection{\textbf{Impact on Tailpipe Emissions}}Fig.~\ref{fig:Experimental_ID_Vrf_Data}-$a$ compares the cumulative tailpipe emissions of the speed and power split optimization scenarios against the baseline case at $T_{cat,0}=50^oC$. The cumulative $\mathcal{Z}$ emission improvement index used in Fig.~\ref{fig:Experimental_ID_Vrf_Data}, where $\mathcal{Z}=~\{HC,CO,NOx\}$, is calculated as follows: \vspace{-0.3cm}
\begin{gather}
    \label{eqn:emission_index}
    \text{Cumulative $\mathcal{Z}$ Improvement}=\frac{\mathcal{Z}^{b}-\mathcal{Z}^*}{\mathcal{Z}^b}\times100
\end{gather}
where $\mathcal{Z}^{b}$ is the emission level resulted from the baseline case, and $\mathcal{Z}^{*}$ represents the eco-driving cases with rule-based (red) and DP-based (green) power split controllers. While the speed optimization (Stage I) on average reduces the CO and NOx emissions by 3.9\% and 7.5\%, respectively, it increases the average HC emissions slightly by 0.1\%. When the power split is also being optimized (Stage I+II), the average CO and NOx emissions are further reduced by 8.8\% and 17.4\%, respectively. However, Fig.~\ref{fig:Experimental_ID_Vrf_Data}-$a_1$ shows that speed and power split optimization lead to an average increase of 3.4\% in HC emissions as compared to the baseline case. Overall, it is observed that sequential optimization not only improves the fuel economy, but also reduces the CO and NOx emissions. The main challenge seems to be the HC emissions which increase as a result of speed and power split optimization.
\begin{figure}[h!]
	\begin{center}
		\includegraphics[width=0.96\columnwidth]{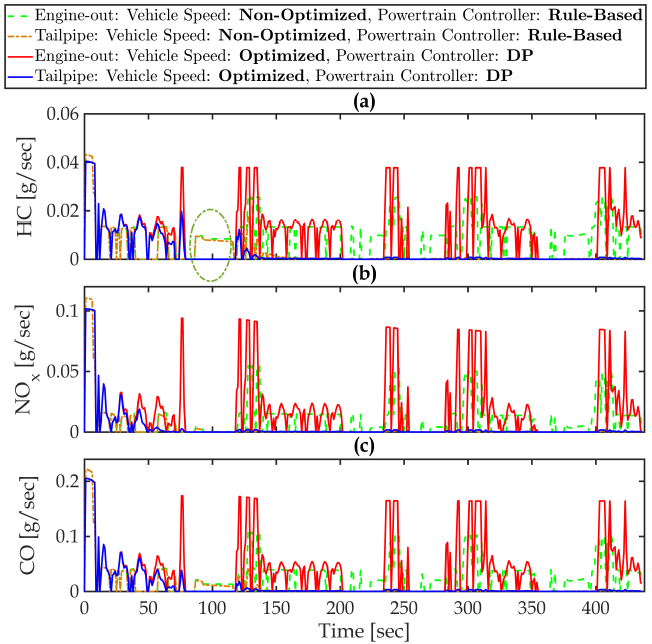} \vspace{-0.45cm}   
		\caption{The results of vehicle speed and power split optimization and its effects on engine-out and tailpipe gas flow rates for the speed profile shown in Fig.~\ref{fig:DP_SpeedOptimization_SOCEng_Nov28_2}: (a) HC, (b) NOx, and (c) CO.}\vspace{-0.85cm} \label{fig:Emissions_RB_noCAV_CAV} 
	\end{center}
\end{figure}

In order to explain the increase in the HC emissions in Fig.~\ref{fig:Experimental_ID_Vrf_Data}-$a_1$, the optimized and non-optimized speed profiles shown in Fig.~\ref{fig:DP_SpeedOptimization_SOCEng_Nov28_2} are recalled, and the flow rates of engine-out and tailpipe HC, NOx, and CO are plotted in Fig.~\ref{fig:Emissions_RB_noCAV_CAV} for the case of rule-based and DP-based power split controllers. As shown in Fig.~\ref{fig:DP_SpeedOptimization_SOCEng_Nov28_2}, the DP may run the engine during the stops. As a result, during the highlighted first long vehicle stop, while with the rule-based controller no HC emissions are generated (engine is off), the engine generates HC when DP is used. At the same time, since the engine is being used less often and more smoothly with the optimized speed and DP-based controller, the TWC light-off occurs later, after the vehicle passes the first stop. On the other hand, with the non-optimized speed profile, the engine is operating more aggressively, resulting in higher heat generation and faster TWC light-off. Note that, according to Fig.~\ref{fig:ATS_Eff_Curves}, the TWC light-off temperature for HC is higher than for CO and NOx. Additionally, Fig.~\ref{fig:Emissions_RB_noCAV_CAV} shows that, compared to CO and NOx, the ``rate'' of the change in engine-out HC emissions flow rate, when the engine is turned on, is higher. 
%

The simulations are repeated at higher initial catalyst temperature, e.g., $T_{cat,0}=300^oC$, and the results are summarized in Fig.~\ref{fig:Experimental_ID_Vrf_Data}-$b$. For CO and NOx, similar trends are observed, where speed optimization reduces the CO and NOx emissions by 14.2\% and 15.3\%, respectively, on average. These numbers are further improved when the power split is optimized as well by 16.2\% (CO) and 26.5\% (NOx). It is noted that with $T_{cat,0}=300^oC$, the TWC is operating ideally with conversion efficiency of $>98\%$ according to Fig.~\ref{fig:ATS_Eff_Curves}. This means that the considerable reduction in the CO and NOx emissions are due to smoother operation of the engine, thanks to the optimization of the speed and power split. Unlike Fig.~\ref{fig:Experimental_ID_Vrf_Data}-$a_1$, Fig.~\ref{fig:Experimental_ID_Vrf_Data}-$b_1$ shows when $T_{cat,0}$ is higher than the light-off temperature, the HC emissions are also reduced by 12.4\% when the speed is optimized. However, according to the earlier argument, when the DP is used for power split strategy, Fig.~\ref{fig:Experimental_ID_Vrf_Data}-$b_3$ shows the HC emissions are only reduced by 4\% on average.

Based on the above results, the overall conclusions are as follows:
\begin{itemize}
    \item eco-driving without specific TWC light-off strategy may delay the TWC light-off and increase the tailpipe HC.
    \item DP-based power split approach may increase engine-out emissions during the vehicle stops.
    \item if the TWC is warmed up and has reached the light-off temperature, eco-driving and power split optimization reduce the tailpipe emissions. 
    \item the trade-off between fuel economy and tailpipe emissions calls for an integrated power and thermal management optimization responding to the fast TWC light-off design requirement for CAVs.
\end{itemize}

\vspace{-0.25cm}
\section{Summary and Conclusions}\label{sec:5}
A sequential optimization framework for eco-driving speed trajectory planning and power split control in hybrid electric CAVs was proposed in this paper, and its impact on the vehicle thermal responses and emission levels was investigated. Simulation results of real-world driving cycles showed that average energy savings of 13\% and 16\% can be achieved sequentially through eco-driving and power split optimization, respectively. Our investigation showed that, due to smoother engine operation, sequential optimization can lead to lower engine and TWC temperatures, which could result in delayed light-off of the TWC. Consequently, the tailpipe HC emissions may be affected unfavorably, and can increase by 3.4\% on average as a result of sequential optimization. This calls for an integrated power split, engine thermal management, and aftertreatment optimization approach to address the trade-off between the fuel economy and optimal vehicle thermal responses.

\vspace{-0.15cm}


\bibliographystyle{unsrt} 
\bibliography{ACC2018Ref.bib} 



\end{document}